\documentclass[aps,onecolumn,amsmath]{revtex4}\textwidth 15cm

\usepackage{amsmath}
\usepackage{graphicx}
\usepackage{amssymb}
\usepackage{epsfig}
\usepackage{amsfonts}

\def \etal{{\em et al.}}

\def\fra#1#2{{\textstyle\frac{#1}{#2}\,}}

\newtheorem{problem}{Problem}

\begin{document}

\title{Paradoxes of measures of quantum entanglement and Bell's
inequality violation in two-qubit systems}

\author{Adam Miranowicz, Bohdan Horst, and Andrzej Koper}
\date{\today}

\affiliation{Faculty of Physics, Adam
 Mickiewicz University, 61-614 Pozna\'n, Poland}

\begin{abstract}
We review some counterintuitive properties of standard measures
describing quantum entanglement and violation of Bell's inequality
(often referred to as ``nonlocality'') in two-qubit systems. By
comparing the nonlocality, negativity, concurrence, and relative
entropy of entanglement, we show: (i) ambiguity in ordering states
with the entanglement measures, (ii) ambiguity of robustness of
entanglement in lossy systems and (iii) existence of two-qubit
mixed states more entangled than pure states having the same
negativity or nonlocality. To support our conclusions, we
performed a Monte Carlo simulation of $10^6$ two-qubit states and
calculated all the entanglement measures for them. Our
demonstration of the relativity of entanglement measures implies
also how desirable is to properly use an operationally-defined
entanglement measure rather than to apply formally-defined
standard measures. In fact, the problem of estimating the degree
of entanglement of a bipartite system cannot be analyzed
separately from the measurement process that changes the system
and from the intended application of the generated entanglement.
\end{abstract}

\maketitle \pagenumbering{arabic}
\section{Introduction}

Quantum entanglement~\cite{Sch35,Ein35}, being at heart of Bell's
theorem~\cite{Bell}, is considered to be an essential resource for
quantum engineering, quantum communication, quantum computation,
and quantum information~\cite{Nielsen}. There were proposed
various entanglement measures and criteria to detect entanglement.
Nevertheless, despite the impressive progress in understanding
this phenomenon (see a recent comprehensive review by Horodecki
\etal~\cite{Horodecki-review} and references therein), a complete
theory of quantum entanglement has not been developed yet.

It is a commonly accepted fact that the entropy of entanglement of
two systems, which is defined to be the von Neumann entropy of one
of the systems, is the unique entanglement measure for bipartite
systems in a pure state~\cite{popescu}. However, in the case of
two systems in a mixed state, there is no unique entanglement
measure. In order to describe properties of quantum entanglement
of bipartite systems various measures have been proposed. Examples
include~\cite{Horodecki-review}: entanglement of formation,
distillable entanglement, entanglement cost, PPT entanglement
cost, the relative entropy of entanglement, or geometrical
measures of entanglement.

It should be stressed that classification of entanglement measures
of mixed states and effective methods of calculation of such
measures are among the most important but still underdeveloped
(with a few exceptions) problems of quantum
information~\cite{open-problems}.

Here, we shortly review counterintuitive properties of some
entanglement measures in the simplest non-trivial case of
entanglement of two qubits.

\section{Measures of quantum entanglement}

We will study quantum entanglement and closely related violation
of Bell's inequality for two qubits in mixed states according to
some standard measures:

(i) To describe the entanglement of formation~\cite{Bennett96} of
a given two-qubit state $\hat\rho$, we apply the Wootters
concurrence~\cite{Wootters} defined as
\begin{equation}
C({\hat\rho})=\max \left(0,2\max_i\lambda_i-\sum_i\lambda_i\right)
\label{N03}
\end{equation}
in terms of $\lambda _{i}$'s, which are the square roots of the
eigenvalues of ${\hat\rho }({\hat\sigma }_2\otimes {\hat\sigma
}_2)\hat\rho^*({ \hat\sigma }_2\otimes {\hat\sigma }_2)$, where
${\sigma }_2$ is the Pauli spin matrix and asterisk stands for
complex conjugation. The concurrence $C({\hat\rho})$ is related to
the entanglement of formation, $E_{F}({\hat\rho})$, as follows
~\cite{Wootters}:
\begin{eqnarray}
E_{F}({\hat\rho}) = {\cal W}[C(\hat\rho)],\quad {\rm where}\quad
{\cal W}(x) \equiv h\left(\frac{1}{2}[1+\sqrt{1-x^2}]\right),
\label{N06a}
\end{eqnarray}
and $h(y)=-y\log _{2}y-(1-y)\log _{2}(1-y)$ is binary entropy.

(ii) The PPT entanglement cost, which is the entanglement
cost~\cite{Horodecki-review} under operations preserving the
positivity of the partial transposition (PPT), can be given
as~\cite{Audenaert03,Ishizaka04}:
\begin{equation}
E_{\rm PPT}(\hat\rho)= \lg[N(\hat\rho)+1] \label{N06}
\end{equation}
in terms of the negativity:
\begin{equation}
N(\hat\rho)=2\sum_{j}\max(0,-\mu _{j}).  \label{N05}
\end{equation}
These measures are related to the Peres-Horodecki
criterion~\cite{Peres,Horodecki96}. In Eq.~(\ref{N05}), $\mu _{j}$
are the eigenvalues of the partial transpose $\hat\rho^{\Gamma}$.

(iii) The relative entropy of entanglement
(REE)~\cite{Vedral97a,Vedral98} is a measure of entanglement
corresponding to a ``distance'' of an entangled state from
separable states. Precisely, the REE can be defined as the minimum
of the relative quantum entropy
\begin{eqnarray}
  S(\hat\rho ||\hat\rho_{\rm sep} )={\rm Tr}\,( \hat\rho \lg
\hat\rho -\hat\rho\lg \hat\rho_{\rm sep} ) \label{N02}
\end{eqnarray}
in the set ${\cal D}$ of all separable states $\hat\rho_{\rm
sep}$, i.e.,
\begin{eqnarray}
E_R(\hat\rho)=\min_{\hat\rho_{\rm sep} \in {\cal D}} S(\hat\rho
||\hat\rho_{\rm sep} )\equiv S(\hat\rho ||{\hat\rho}_{\rm css}),
\label{A2}
\end{eqnarray}
where ${\hat\rho}_{\rm css}$ denotes the closest separable state
(CSS) to $\hat\rho$. Numerical problems to calculate the REE are
shortly discussed in Appendix A.

(iv) To describe a degree of violation of Bell's
inequality~\cite{Bell} due to Clauser, Horne, Shimony and Holt
(CHSH)~\cite{Clauser}, we use the modified Horodecki
measure~\cite{Horodecki95,H4}:
\begin{eqnarray}
B(\hat\rho )\equiv \sqrt{ \max
\,[0,\,\max_{j<k}\;(u_{j}+u_{k})-1\,]}, \label{N34}
\end{eqnarray}
which is given in terms of the eigenvalues $u_{j}$ $(\,j=1,2,3)$
of $U_{\hat\rho}=T_{\hat\rho }^{T}\,T_{\hat\rho }$, where
$T_{\hat\rho }$ is a real matrix with elements
$t_{nm}=\mathrm{Tr}\,[\hat\rho \,(\hat\sigma _{n}\otimes
\hat\sigma _{m})]$, $T_{\hat\rho }^{T}$ is the transposition of
$T_{\hat\rho }$ and $\hat\sigma _{n}$ $(\,n=1,2,3)$ are Pauli's
spin matrices. For short, we refer to $B$ as ``nonlocality''
(measure).

For any two-qubit pure state $|\psi\rangle$, the nonlocality $B$
is equal to the entanglement measures $C$ and $N$:
\begin{eqnarray}
B(|\psi\rangle) =C(|\psi\rangle)=N(|\psi\rangle). \label{N35}
\end{eqnarray}
It is seen that for this case the measures $B$, $C$ and $N$
correspond to the relative entropy of entanglement $E_R$ and von
Neumann's entropy:
\begin{eqnarray}
{\cal W}[B(|\psi\rangle)] ={\cal W}[C(|\psi\rangle)] ={\cal
W}[N(|\psi\rangle)]=E_{R}(|\psi\rangle)=E_{\rm Neumann
}(|\psi\rangle), \label{N35b}
\end{eqnarray}
where ${\cal W}$ is given in Eq.~(\ref{N06a}).

In the following we describe somewhat surprising properties of the
entanglement measures for two-qubits in {\em mixed} states. For
brevity, by referring to the entanglement measures, we also mean
the nonlocality $B$.

\section{Ambiguity in ordering states with entanglement measures}

The problem can be posed as follows:
\begin{problem}
Two measures of entanglement, say ${\cal E}'$ and ${\cal E}''$,
imply the same ordering of states if the condition~\cite{Eisert99}
\label{P1}
\begin{eqnarray}
{\cal E}'(\hat\rho_1) < {\cal E}'(\hat\rho_2) \Leftrightarrow
{\cal E}''(\hat\rho_1) < {\cal E}''(\hat\rho_2) \label{N36}
\end{eqnarray}
is satisfied for arbitrary states $\hat\rho_1$ and $\hat\rho_2$.
The question is whether this condition is fulfilled for all
``good'' entanglement measures.
\end{problem}

In early fundamental works on quantum information, it is often
claimed that good entanglement measures should fulfill this
condition. For example, in Ref.~\cite{Vedral97a} it was stated
that: ``For consistency, it is only important that if $\hat\rho_1$
is more entangled then $\hat\rho_2$ for one measure than it also
must be for all other measures.''

For qubits in pure states, condition~(\ref{N36}) is always
fulfilled, since all good measures are equivalent. However,
standard measures can imply different ordering of mixed states
even for only two qubits. This was first shown numerically by
Eisert and Plenio~\cite{Eisert99} by analyzing their results of
Monte Carlo simulations of two-qubit states. The problem was then
analyzed by
others~\cite{Zyczkowski,Virmani,Wei,H4,H5,H6,H7,Ziman,Kinoshita07,H8}.

To our knowledge, the first analytical examples of two-qubit
states violating condition~(\ref{N36}) were given in
Refs.~\cite{H4,H5}. In Ref.~\cite{H6}, to find analytical examples
of extreme violation of Eq.~(\ref{N36}), we applied the results of
Verstraete \etal~\cite{Verstraete01a} concerning allowed values of
the negativity $N$ for a given value of the concurrence $C$.

Note that the violation of condition~(\ref{N36}) cannot be
observed for pure states of two-qubit systems. By contrast, for
three-level systems (the so-called qutrits), analytical examples
of violation of the condition are known even for pure
states~\cite{Zyczkowski,Virmani,Wei}.

The property that ordering of states depends on the applied
entanglement measure sounds counterintuitive. Nevertheless, it is
physically sound, since states, which are differently ordered
according to two measures, cannot be transformed into each other
with 100\% efficiency by applying local quantum operations and
classical communication (LOCC) only. Virmani and
Plenio~\cite{Virmani} proved in general terms that all good
asymptotic entanglement measures are either identical or have to
imply a different ordering on some quantum states.

In Ref.~\cite{H7}, the three measures (the negativity,
concurrence, and the REE) were compared and found analytical
examples of states (say $\hat\rho'$ and $\hat\rho''$) for which
one measure implies state ordering opposite to that implied by the
other two measures:
\begin{align}
&C(\hat\rho')<C(\hat\rho''),\quad
N(\hat\rho')<N(\hat\rho''),\quad
E_R(\hat\rho')>E_R(\hat\rho'');\nonumber  \\
&C(\hat\rho')<C(\hat\rho''),\quad
N(\hat\rho')>N(\hat\rho''),\quad
E_R(\hat\rho')<E_R(\hat\rho'');\nonumber \\
&C(\hat\rho')>C(\hat\rho''),\quad
N(\hat\rho')<N(\hat\rho''),\quad
E_R(\hat\rho')<E_R(\hat\rho'').
\end{align}
There can be found other analytical examples of states exhibiting
even more peculiar ordering of states according to these three
measures. Examples include pairs of states for which a degree of
entanglement is preserved according to one or two measures but it
is different according to the other measures, e.g.:
\begin{align}
 &C(\hat\rho')=C(\hat\rho''), \quad N(\hat\rho')<N(\hat\rho''), \quad E_R(\hat\rho')>E_R(\hat\rho''); \nonumber \\
 &C(\hat\rho')<C(\hat\rho''), \quad N(\hat\rho')=N(\hat\rho''), \quad E_R(\hat\rho')>E_R(\hat\rho''); \nonumber \\
 &C(\hat\rho')<C(\hat\rho''), \quad N(\hat\rho')>N(\hat\rho''), \quad E_R(\hat\rho')=E_R(\hat\rho'').
\end{align}
and
\begin{align}
&C(\hat\rho')=C(\hat\rho''), \quad N(\hat\rho')=N(\hat\rho''), \quad E_R(\hat\rho')<E_R(\hat\rho'');\nonumber \\
&C(\hat\rho')=C(\hat\rho''), \quad N(\hat\rho')<N(\hat\rho''), \quad E_R(\hat\rho')=E_R(\hat\rho'');\nonumber \\
&C(\hat\rho')<C(\hat\rho''), \quad N(\hat\rho')=N(\hat\rho''), \quad E_R(\hat\rho')=E_R(\hat\rho'').
\end{align}

The comparative analyses presented in Refs.~\cite{H4,H5,H6,H7} are
not only related to a mathematical problem of classification of
states according to various entanglement measures. They could also
enable a deeper understanding of some physical aspects of
entanglement.

\subsection{Nonequivalent states with the same entanglement according
to $E_R$, $C$ and $N$}

\begin{problem}
Find analytical examples of *nonequivalent* two-qubit states
$\hat\rho'$ and $\hat\rho''$ exhibiting the same entanglement of
formation [$C(\hat\rho')=C(\hat\rho'')$], the same PPT
entanglement cost [$N(\hat\rho')=N(\hat\rho'')$], and the same
relative entropy of entanglement
[$E_R(\hat\rho')=E_R(\hat\rho'')$]?
\end{problem}

As a first attempt to find such an example, let us compare two
different pure states:
\begin{eqnarray}
|\psi' \rangle &=& c'_{00}|00\rangle +c'_{01}|01\rangle
+c'_{10}|10\rangle +c'_{11}|11\rangle, \nonumber\\
|\psi'' \rangle &=& c''_{00}|00\rangle +c''_{01}|01\rangle
+c''_{10}|10\rangle +c''_{11}|11\rangle, \label{N39}
\end{eqnarray}
fulfilling the condition
\begin{eqnarray}
|c'_{00}c'_{11}-c'_{01}c'_{10}|=|c''_{00}c''_{11}-c''_{01}c''_{10}|,
\end{eqnarray}
which guarantees the same degree of entanglement according to the
measures $C$, $N$ and $E_R$. However, states $|\psi' \rangle$ and
$|\psi'' \rangle$ can be transformed into each other by local
operations. Namely, by applying local rotations, $|\psi \rangle$
can be converted into ($p=p',p''$)
\begin{eqnarray}
|\tilde\psi(p) \rangle &=&\sqrt{p} |01\rangle + \sqrt{1-p}
|10\rangle  \label{N40}
\end{eqnarray}
for which the negativity and concurrence are equal to
$2\sqrt{p(1-p)}$. The same value is obtained also for
$|\tilde{\psi}(1-p) \rangle$, but this state can be transformed
into $|\tilde{\psi}(p) \rangle$ by applying the NOT gate to each
of the qubits. This shows that pure states are {\em not} a good
example of states satisfying the conditions specified in
Problem~2.

As a second attempt, let us compare two Bell diagonal states
described by $\hat\rho'_{B}$ and $\hat\rho''_{B}$ with the same
maximum eigenvalue $\max_i \lambda_i>1/2$. These states have the
same entanglement according to the measures $C$, $N$ and $E_R$.
However, as shown in Ref.~\cite{H7}, states $\hat\rho'_{B}$ and
$\hat\rho''_{B}$ exhibit different nonlocality, i.e., violate
Bell's inequality to different degree. Specifically, the
nonlocality $B$ for a Bell diagonal state is given by~\cite{H7}:
\begin{eqnarray}
B(\hat\rho_B)=\sqrt{\max\{0,2\max_{(i,j,k)}[(\lambda_i-\lambda_j)^2
+(\lambda_k-\lambda_4)^2]-1\}}, \label{N41}
\end{eqnarray}
where subscripts $(i,j,k)$ correspond to cyclic permutations of
$(1,2,3)$. It is seen that violation of Bell's inequality depends
on all values of $\lambda_i$, while the entanglement measures
$E_R$, $C$, and $N$ depend only on the largest value $\max_i
\lambda_i>1/2$. Thus, states $\hat\rho'_{B}$ and $\hat\rho''_{B}$,
fulfilling the conditions ${\rm eig}(\hat\rho'_{B})\neq {\rm
eig}(\hat\rho''_{B})$ and $\max\{{\rm eig}(\hat\rho'_{B})\}=\max
\{{\rm eig}(\hat\rho''_{B})\}>1/2$, have the same entanglement
measures: $E_R(\hat\rho'_{B})=E_R(\hat\rho''_{B})$,
$C(\hat\rho'_{B})=C(\hat\rho''_{B})$ and
$N(\hat\rho'_{B})=N(\hat\rho''_{B})$, but the states are not
equivalent as they exhibit different nonlocality,
$B(\hat\rho'_{B})\neq B(\hat\rho''_{B})$.

\section{Ambiguity of robustness of entanglement}

\subsection{Maximally entangled pure states in lossy cavities}

Let us analyze the following problem:
\begin{problem}
Which maximally entangled pure states are the most fragile or
robust to decoherence of two qubits in lossy cavities?
\end{problem}
\begin{figure}
\hspace*{2mm}
\includegraphics[width=4.25cm]{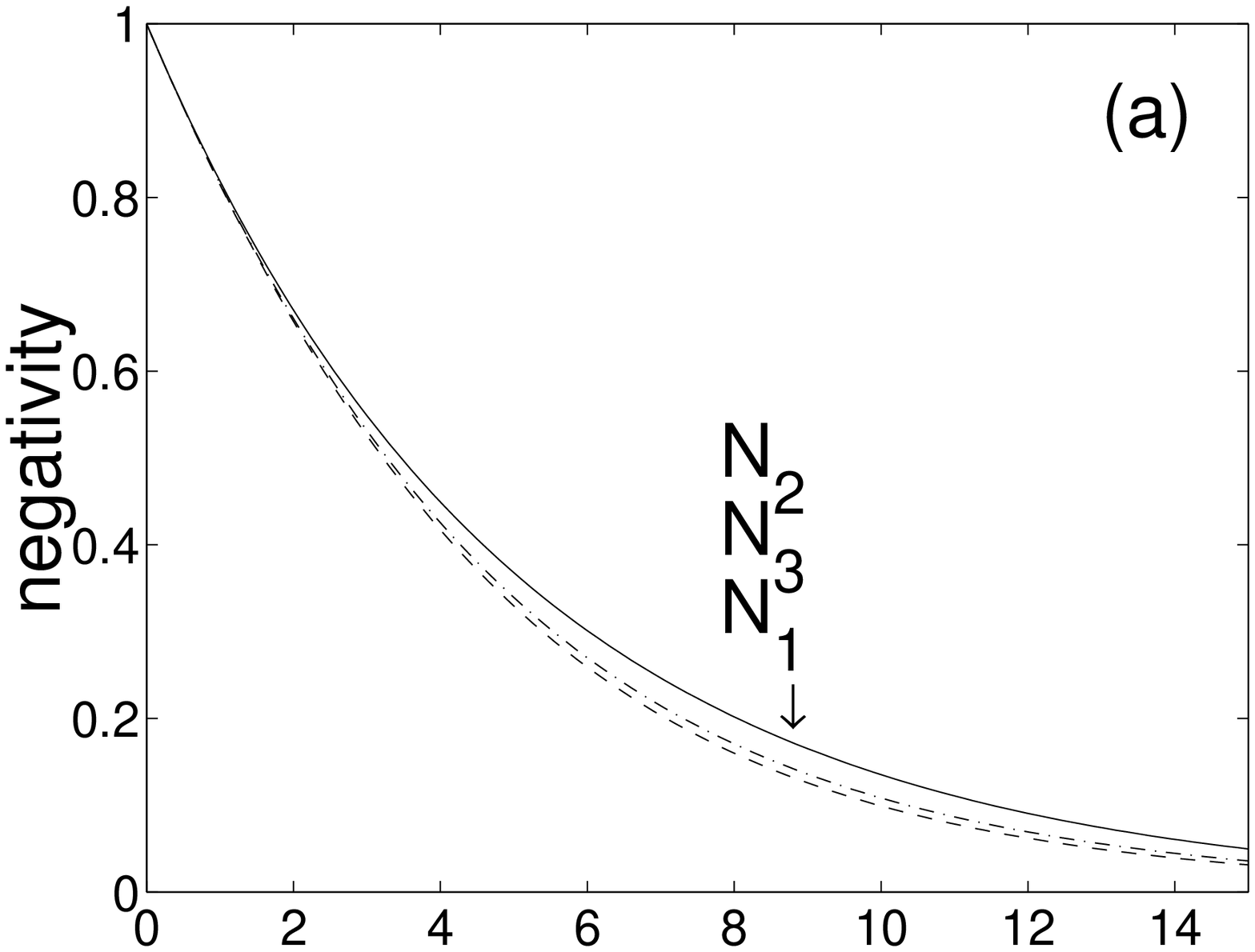}
\hspace*{0cm}
\includegraphics[width=4.5cm]{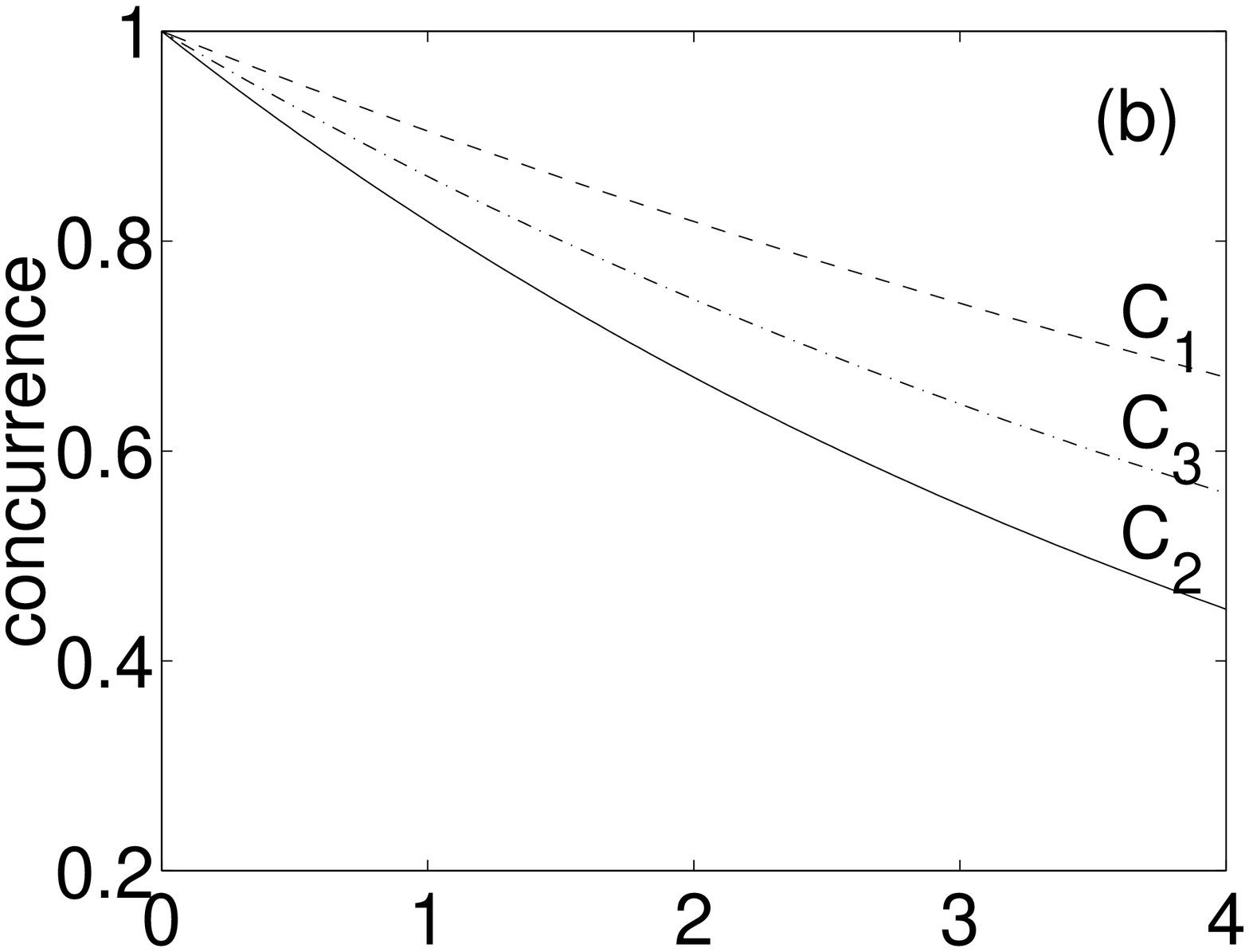}
\hspace*{0cm}
\includegraphics[width=4.5cm]{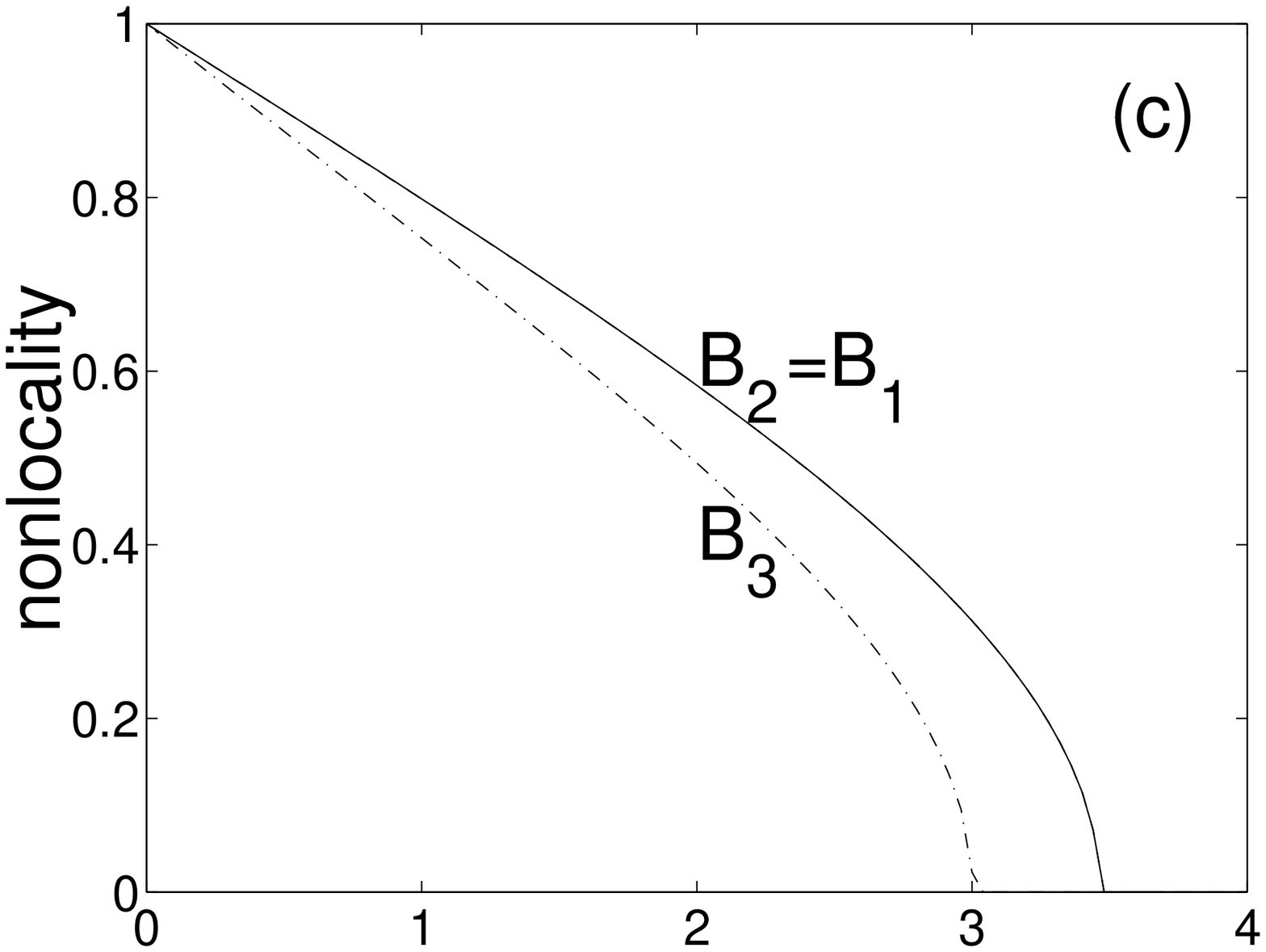}

\vspace*{-1mm}{\hspace{7mm} rescaled time\hspace{28mm}rescaled
time\hspace{28mm}rescaled time}

\vspace*{0cm} \caption{Decay of entanglement between two qubits
initially in the maximally entangled states $|\Psi_k\rangle$ (for
$k=1,2,3$) in lossy cavities with damping rates $\gamma=0.1$
described by: (a) the negativity $N$, (b) the concurrence $C$, and
(c) the nonlocality $B$. It is seen that there is no simple answer
to the question which of the initial states $|\Psi_k\rangle$ is
the most fragile (or robust) to decoherence. In the discussed
model of dissipation, the fastest decoherence exhibits:
$|\Psi_1\rangle$ according to $N$, $|\Psi_2\rangle$ according to
$C$,  and $|\Psi_3\rangle$ according to $B$.}
\end{figure}
This problem was addressed in Refs.~\cite{H4,H5} by analyzing
decoherence of optical photon-number qubits stored initially in
the following three maximally entangled (pure) states (MES):
\begin{eqnarray}
|\Psi_1\rangle =\frac{1}{\sqrt{2}}(|01\rangle - |10\rangle), \quad
|\Psi_2\rangle=\frac{1}{\sqrt{2}} (|00\rangle + |11\rangle),
\label{x2}
\end{eqnarray}
\begin{eqnarray}
|\Psi_3\rangle =\frac{1}{2}(|00\rangle +|01\rangle +|10\rangle
-|11\rangle )\equiv \frac{1}{\sqrt{2}}(|0,+\rangle +|1,-\rangle ),
\label{x3}
\end{eqnarray}
where $|\pm \rangle =(|0\rangle \pm |1\rangle )/\sqrt{2}$. State
$|\Psi_3\rangle$ can be obtained from $|\Psi_2\rangle$ by applying
Hadamard's gate to the second qubit.

To address Problem 3, let us analyze two entangled qubits in a
superposition of vacuum and single-photon states (so-called
photon-number qubits) in a lossy cavity (or, equivalently, in two
cavities). Then, one can apply the standard master-equation
approach to describe the effect of radiative decay of cavities
(i.e., zero-temperature reservoirs) on entanglement of two qubits
according to the concurrence $C_{k}(t)$, negativity $N_{k}(t)$,
and nonlocality $B_{k}(t)$~\cite{H4}. In Fig. 1, it is assumed
that the qubits are initially in the MES $|\Psi_k\rangle$ for
$k=1,2,3$ and the cavity damping rate is $\gamma=0.1$. By
analyzing Fig.~1, one can conclude that entanglement decays in
this model fulfill the inequalities:
\begin{eqnarray}
N_{2}(t) &\ge&  N_{3}(t) \;\ge\; N_{1}(t),\nonumber \\
B_{1}(t) &=&  B_{2}(t) \;\ge\; B_{3}(t), \nonumber \\
C_{1}(t) &\ge&  C_{3}(t) \;\ge\; C_{2}(t).
\end{eqnarray}
It is worth noting that due to the Markov approximation assumed in
the derivation of the master equation, our conclusions are valid
for evolution times $t$ short in comparison to reservoir decay
time $\gamma^{-1}$, and much longer than correlation time $\tau_c$
of reservoir(s), i.e., $\tau_c \ll t-t_0\ll \gamma_j^{-1}$, where
$t_0$ is the initial evolution time. Thus, in this specific
dissipation model, the most fragile to dissipation is
$|{\Psi}_{1}\rangle$ according to the negativity $N$,
$|{\Psi}_{2}\rangle$ according to the concurrence $C$, and
$|{\Psi}_{3}\rangle$ according to the nonlocality $B$. The results
seem to be contradicting, but it should be remembered that
measures $C$, $N$ and $B$ describe different aspects of mixed
states even if for pure states they coincide $C=N=B$. Results of
Refs.~\cite{H4,H5} clearly confirm the relativity of state
ordering by $C$, $N$ and $B$. This example of Ref.~\cite{H4} was
probably the first demonstration of this property in a real
physical process.

\subsection{Maximally entangled mixed states in lossy cavities}

Here, we analyze decay of Werner's states, which can be defined
for $p\in \langle 0,1\rangle$ as~\cite{Werner89}:
\begin{eqnarray}
{\hat\rho }_{1}^{(p)}(0) &=& p|\Psi_1\rangle \langle \Psi_1|
+\fra{1-p}{4}\hat I\otimes {\hat I}, \label{N46}
\end{eqnarray}
which is a mixture of the singlet state, $|\Psi_1\rangle$, and
maximally mixed state, given by $\hat I\otimes \hat I$, where
$\hat I$ is identity operator. Original Werner's state can be
generalized for mixtures of other Bell states with $\hat I\otimes
\hat I$. Thus, one can define Werner-type state as follows
($k=2,3$):
\begin{eqnarray}
{\hat\rho }_{k}^{(p)}(0) &=&p|\Psi_k\rangle \langle \Psi_k|+
\fra{1-p}{4}\hat I\otimes \hat I, \label{N47}
\end{eqnarray}
where $|\Psi_2\rangle$ and $|\Psi_3\rangle$ are given by
Eqs.~(\ref{x2}) and~(\ref{x3}), respectively.

Werner's states can be considered as {\em maximally entangled
mixed states} (MEMS) of two qubits since the amount of
entanglement of these states cannot be increased by any unitary
transformation~\cite{ishizaka00} and they are maximally entangled
(according to the concurrence) for a given value of linear
entropy~\cite{munro01}.

Let us ask more specific question related to Problem~3:
\begin{problem}
Which MEMS are the most robust to dissipation
in the discussed model of lossy cavities?
\end{problem}
\begin{figure}
\centerline{
\includegraphics[width=6cm]{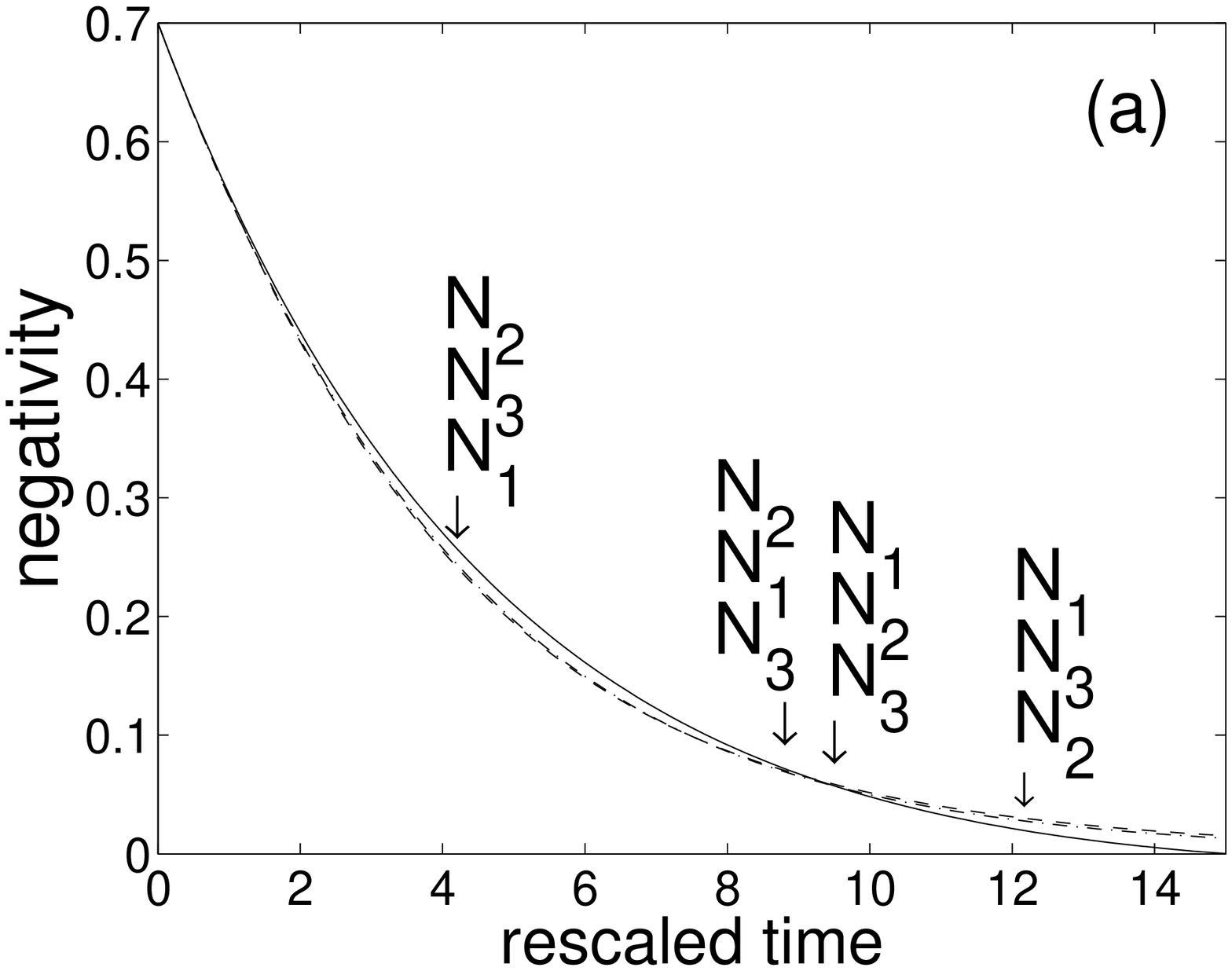}
\hspace*{0cm}
\includegraphics[width=6cm]{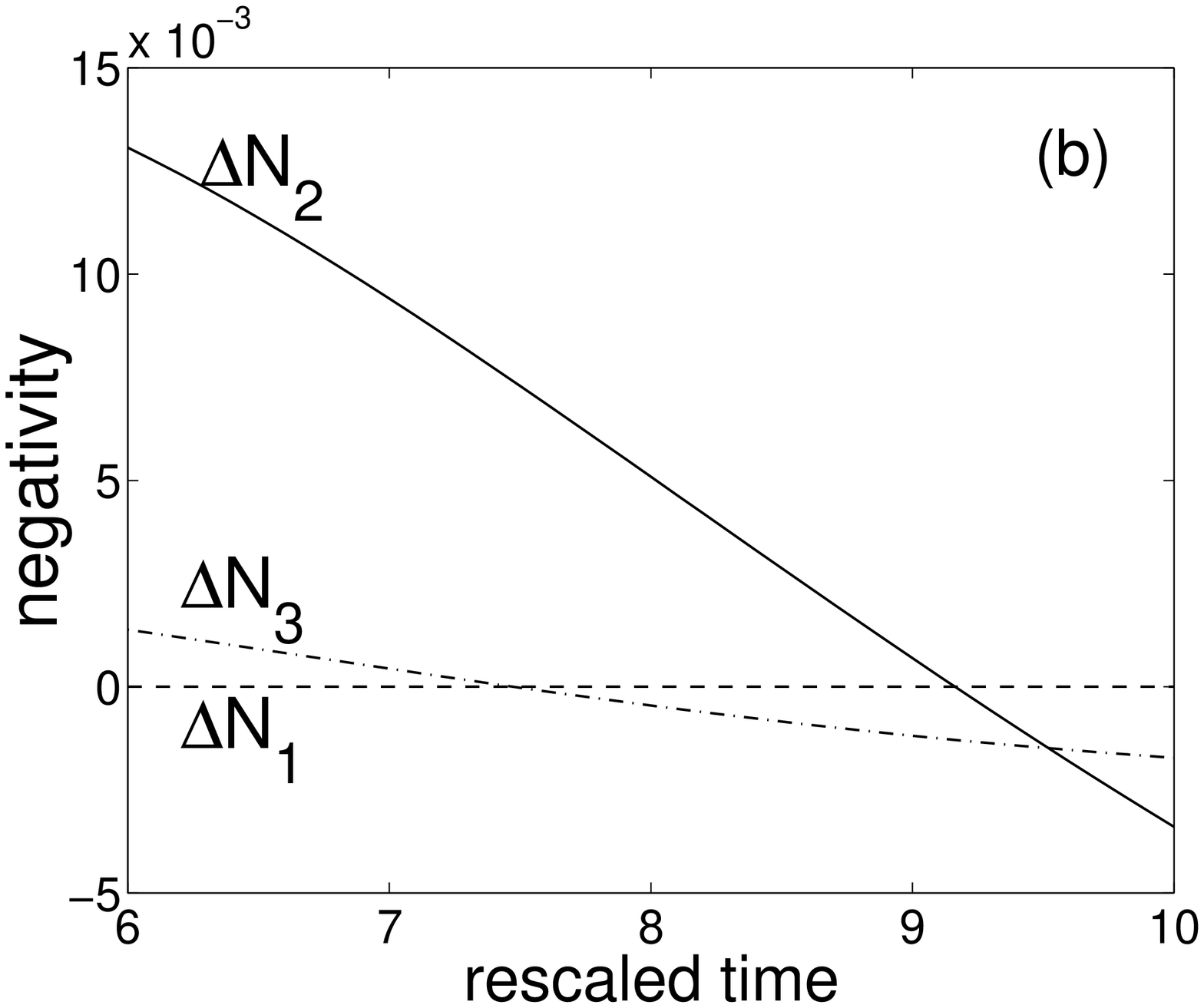}}
\vspace*{0cm} \caption{Decay of entanglement of two photon-number
qubits in a lossy cavity. Entanglement is measured by the
negativity (a) $N_k$ and (b) $\Delta N_{k}= N_{k}- N_{1}$ for
qubits initially in Werner's states ${\hat\rho }_{k}^{(p)}(0)$ for
$k=1,2,3$ and $p=0.8$. The cavities damping rate is $\gamma=0.1$.
For clarity, the scale of figure (b) is enlarged in comparison to
figure (a).}
\end{figure}

Even for such formulated question there is no simple answer. To
show this we analyze the same model of decaying photon-number
qubits in a lossy cavity (or cavities) as studied in Sect.~4.A,
but for qubits initially in Werner's states ${\hat\rho
}_{k}^{(p)}(0)$ for $k=1,2,3$ and $p=0.8$. Let us compare the
decays of the negativity as shown in Fig. 2 and also described in
detail in Table I in Ref.~\cite{H4}. It is seen that a given
Werner state can be more robust to decay than another Werner's
state at short evolution times but, in turn, less robust at longer
times. The differences between the negativity values for various
states shown in Fig.~2 are not very large but still distinct.

\section{Mixed states more entangled than pure states}

\begin{problem}
Can two-qubit *mixed* states be more entangled than *pure* states
according to some entanglement measure ${\cal E}'$ at a fixed
value of another entanglement measure ${\cal E}''$ assuming ${\cal
E}'(\hat{\rho})\le {\cal E}''(\hat{\rho})$ for any state
$\hat{\rho}$?
\end{problem}

\begin{figure}
\centerline{
\includegraphics[width=5cm]{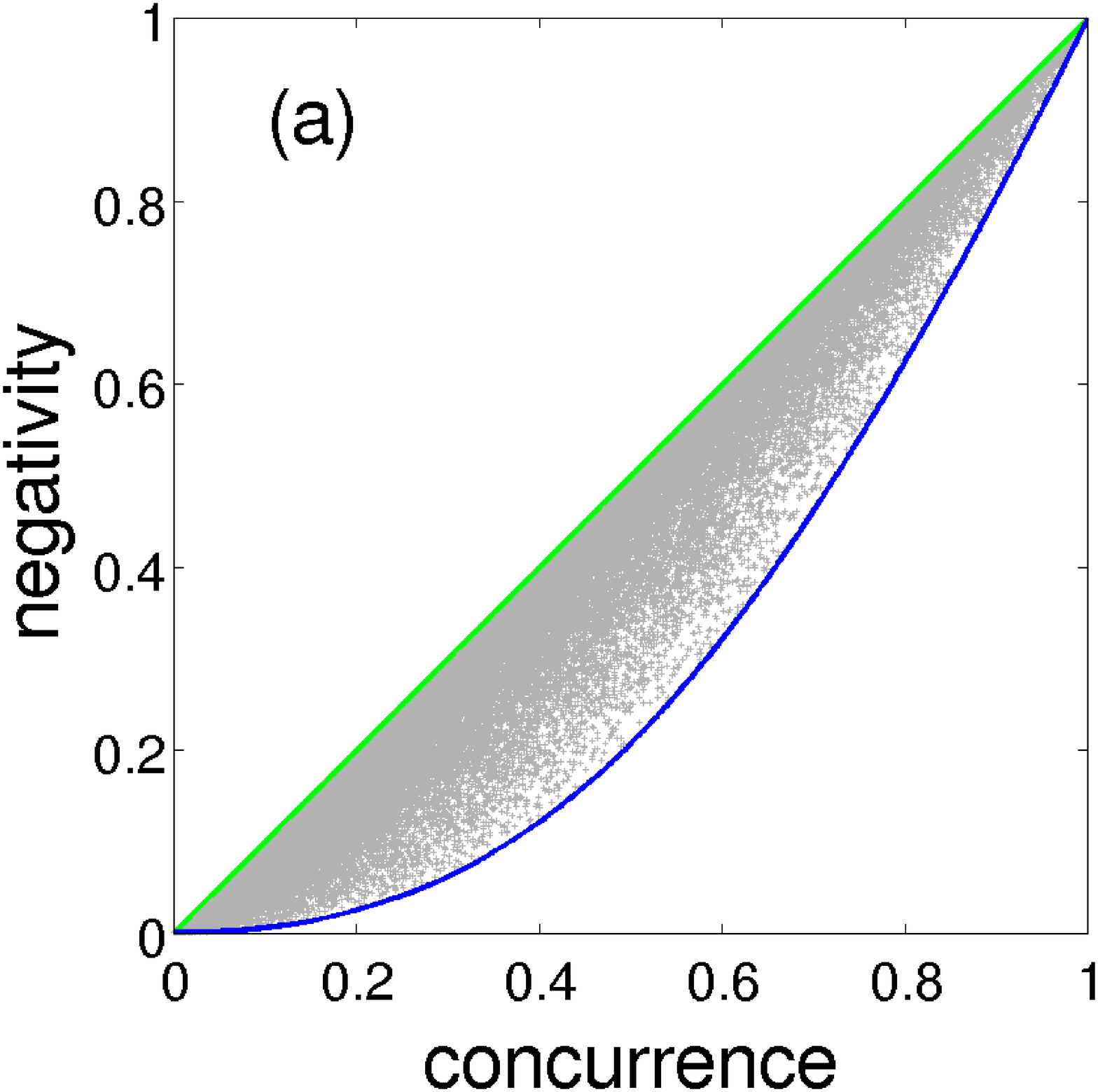}
\hspace*{0cm}
\includegraphics[width=5cm]{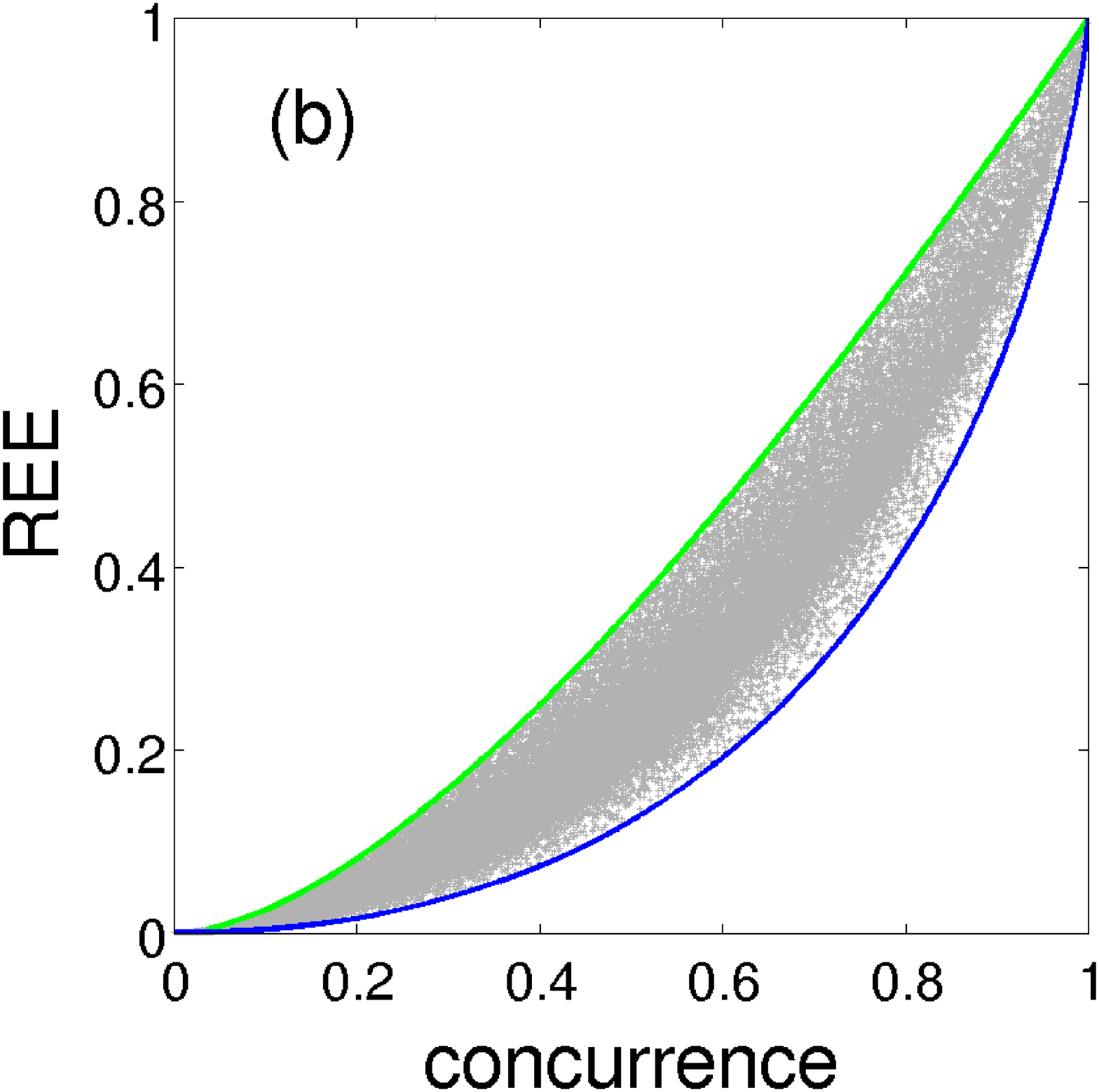}
\hspace*{0cm}
\includegraphics[width=5cm]{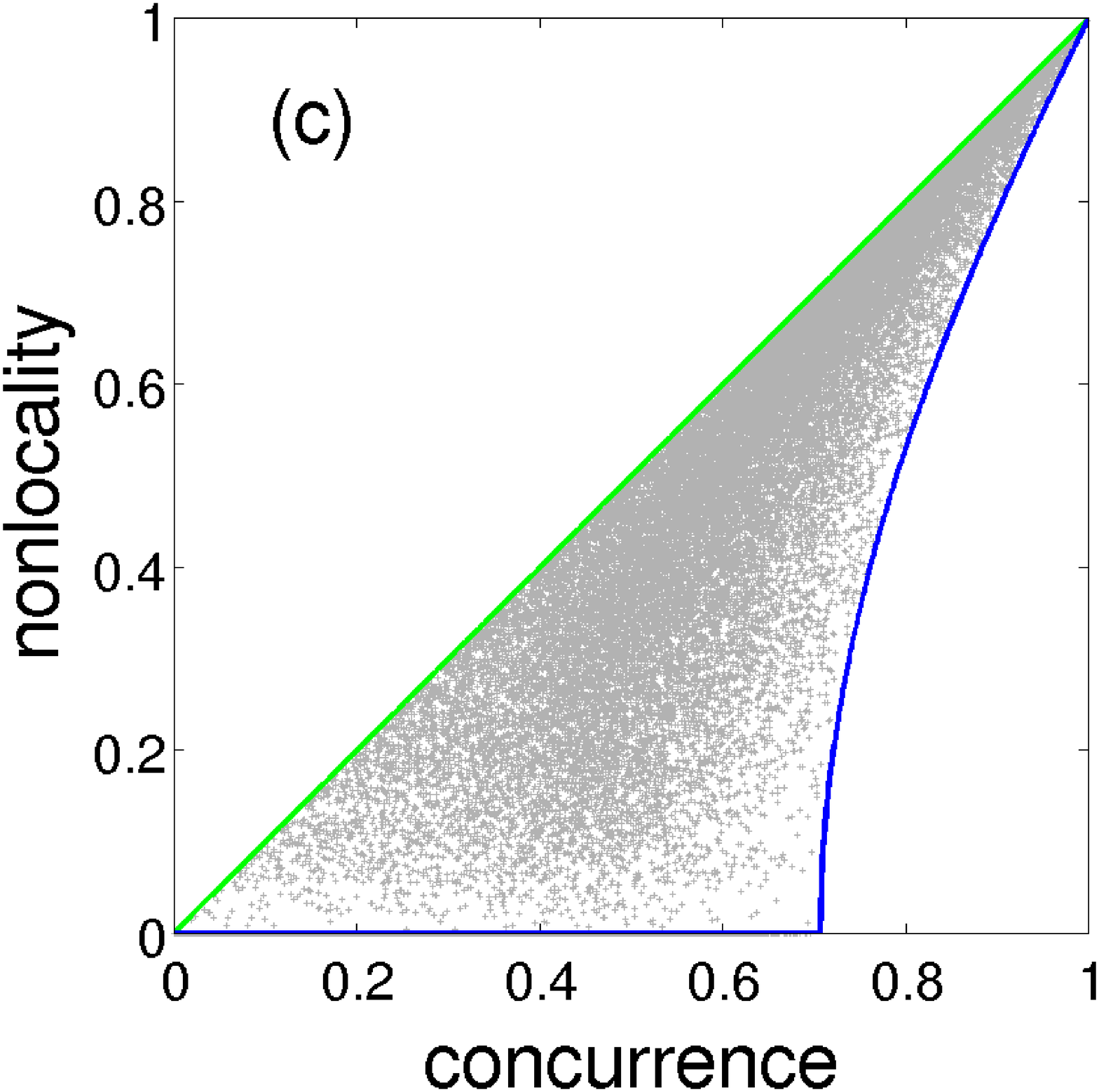}}

\centerline{
\includegraphics[width=5cm]{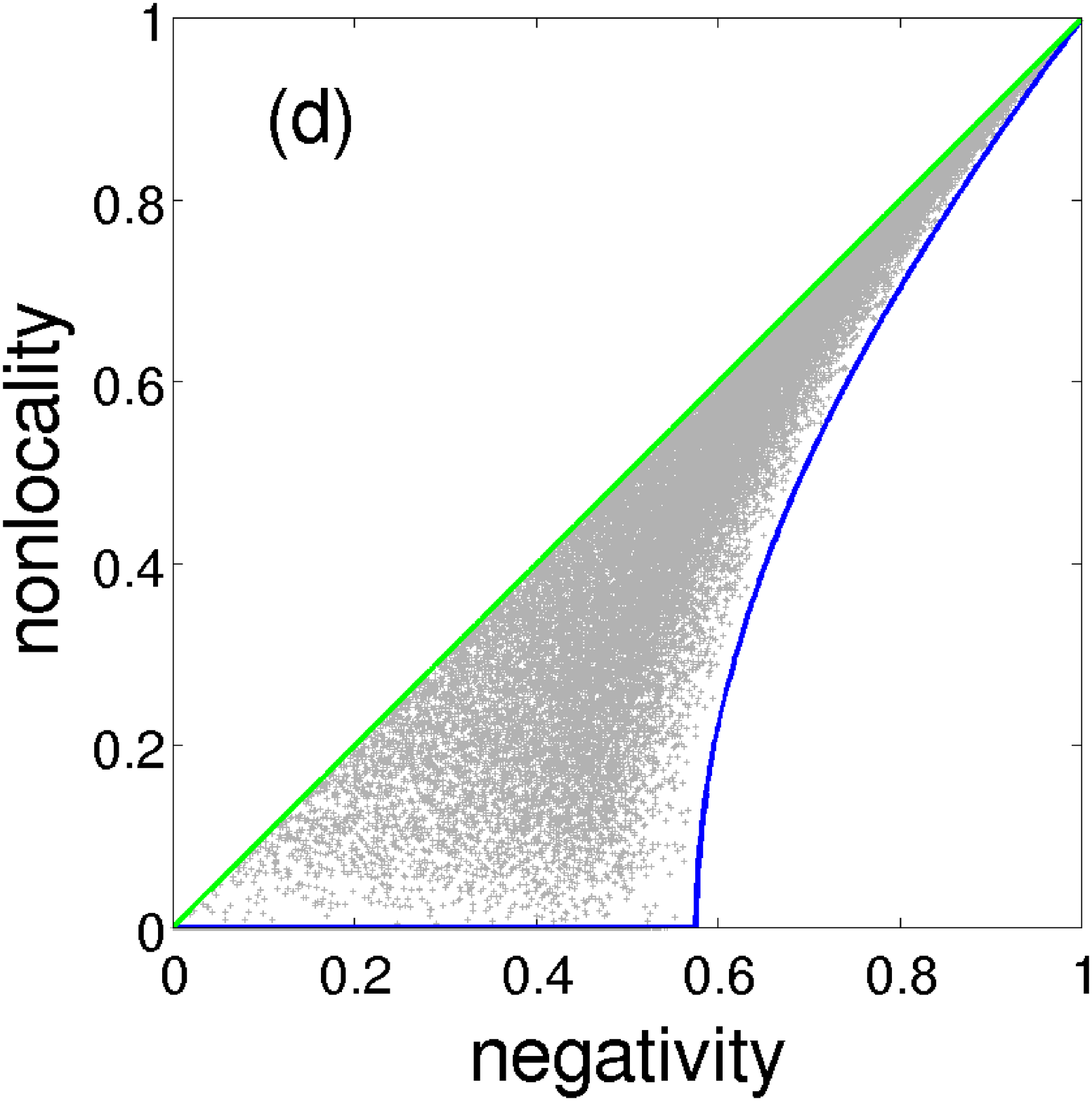}
\hspace*{0cm}
\includegraphics[width=5cm]{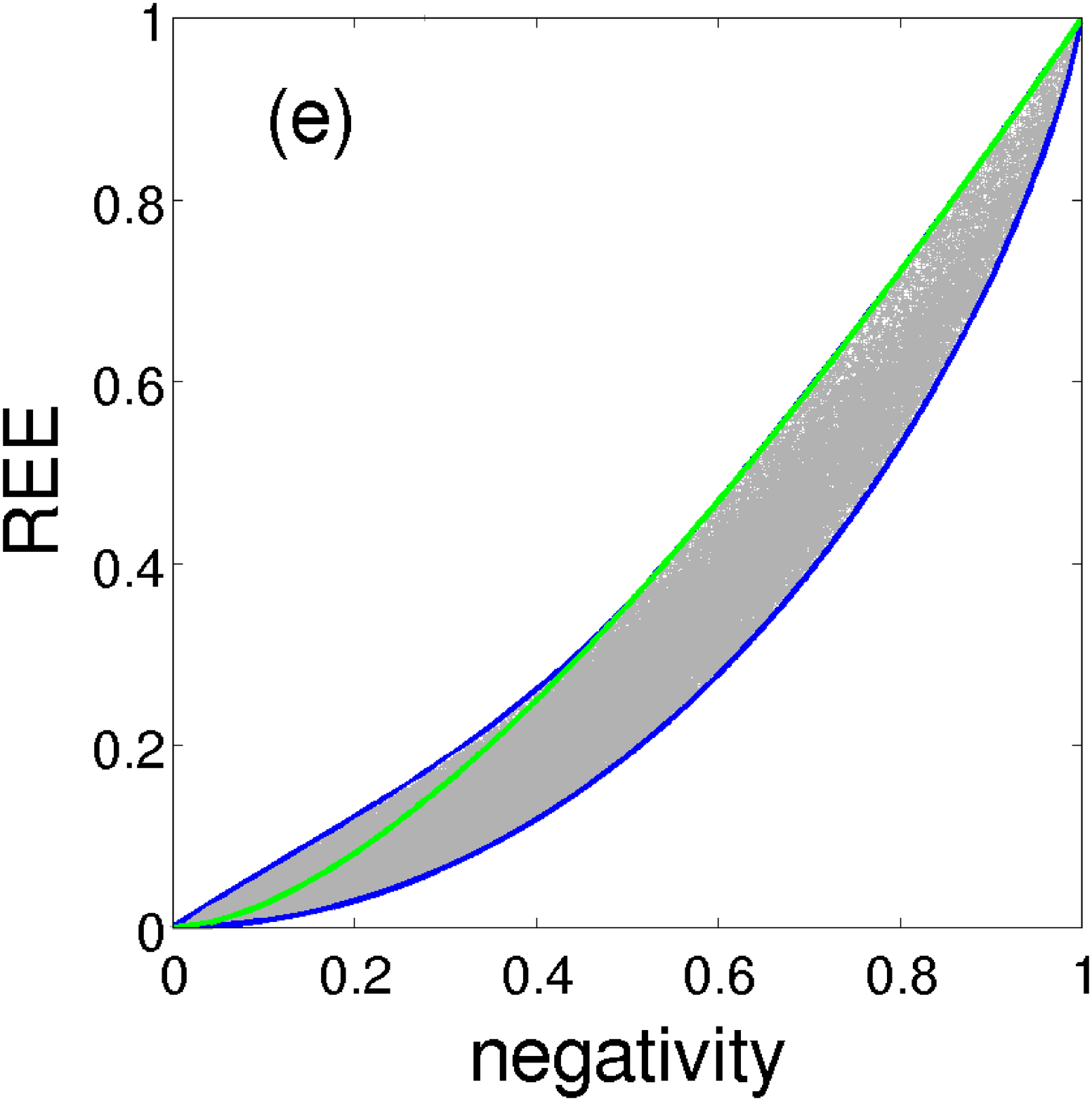}
\hspace*{0cm}
\includegraphics[width=5cm]{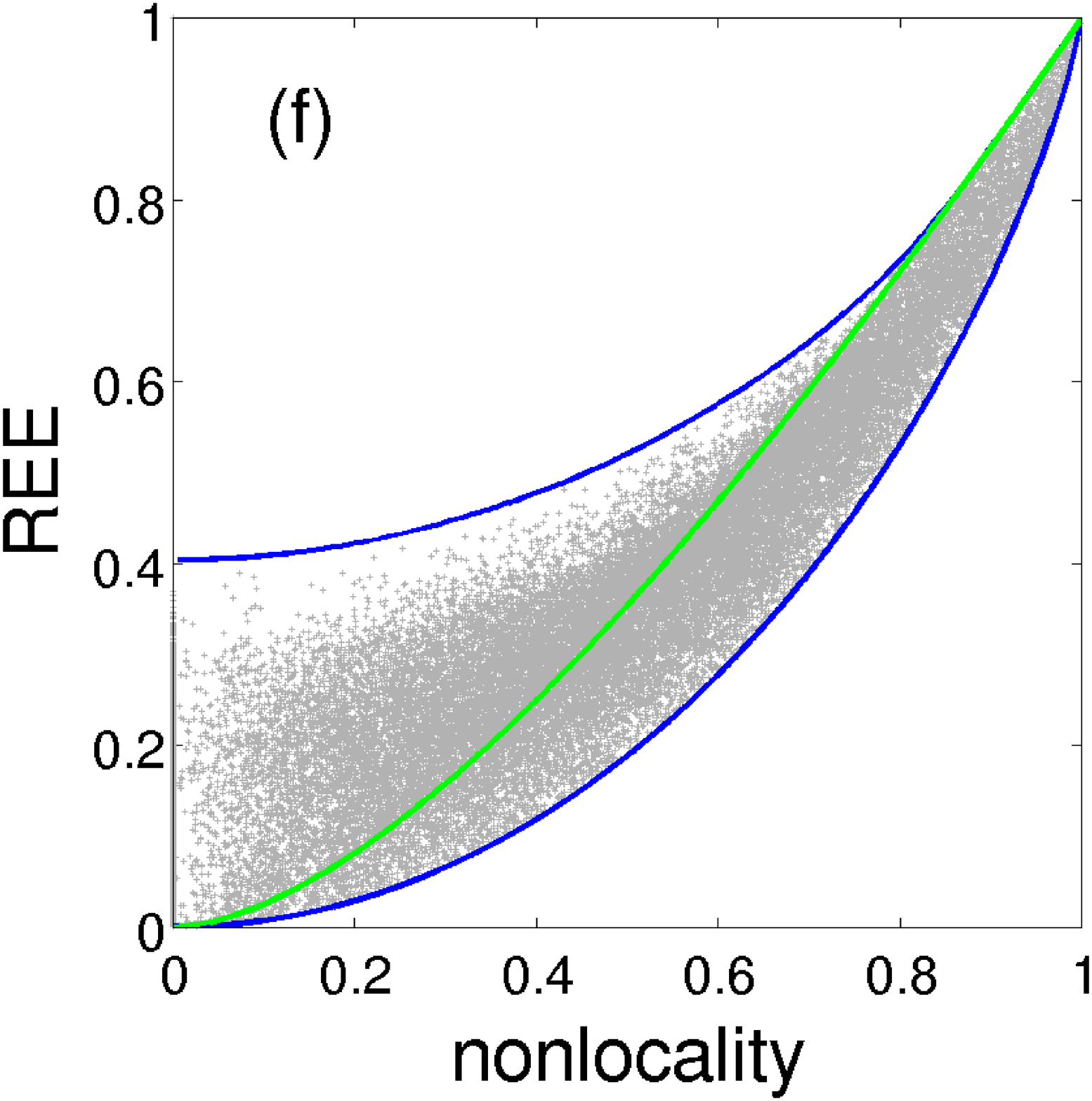}}
\vspace*{0cm} \caption{(Color online) Entanglement and nonlocality
measures for $10^6$ two-qubit states $\hat\rho$ generated by a
Monte Carlo simulation. Green curves correspond to pure states,
and blue curves show the upper and lower bounds of a one measure
${\cal E}'(\hat\rho)$ as a function of another ${\cal
E}''(\hat\rho)$. It is seen that pure states $|\psi\rangle$ lie
for the whole range of abscissa at the upper bound of: (a) the
negativity $N(\hat\rho)$ for a given value of the concurrence
$C(\hat\rho)$ (b) the REE $E_R(\hat\rho)$ vs $C(\hat\rho)$, (c)
the nonlocality $B(\hat\rho)$ vs $C(\hat\rho)$, and (d)
$B(\hat\rho)$ vs $N(\hat\rho)$. However, for (e) $E_R(\hat\rho)$
vs $N(\hat\rho)$ and (e) $E_R(\hat\rho)$ vs $B(\hat\rho)$ pure
states are at the upper bound for abscissa values close to one
only. Thus, in the cases (e) and (f), the entanglement of mixed
states can exceed that of pure states for abscissa values close to
zero.}
\end{figure}
\begin{figure}
\centerline{
\includegraphics[width=5cm]{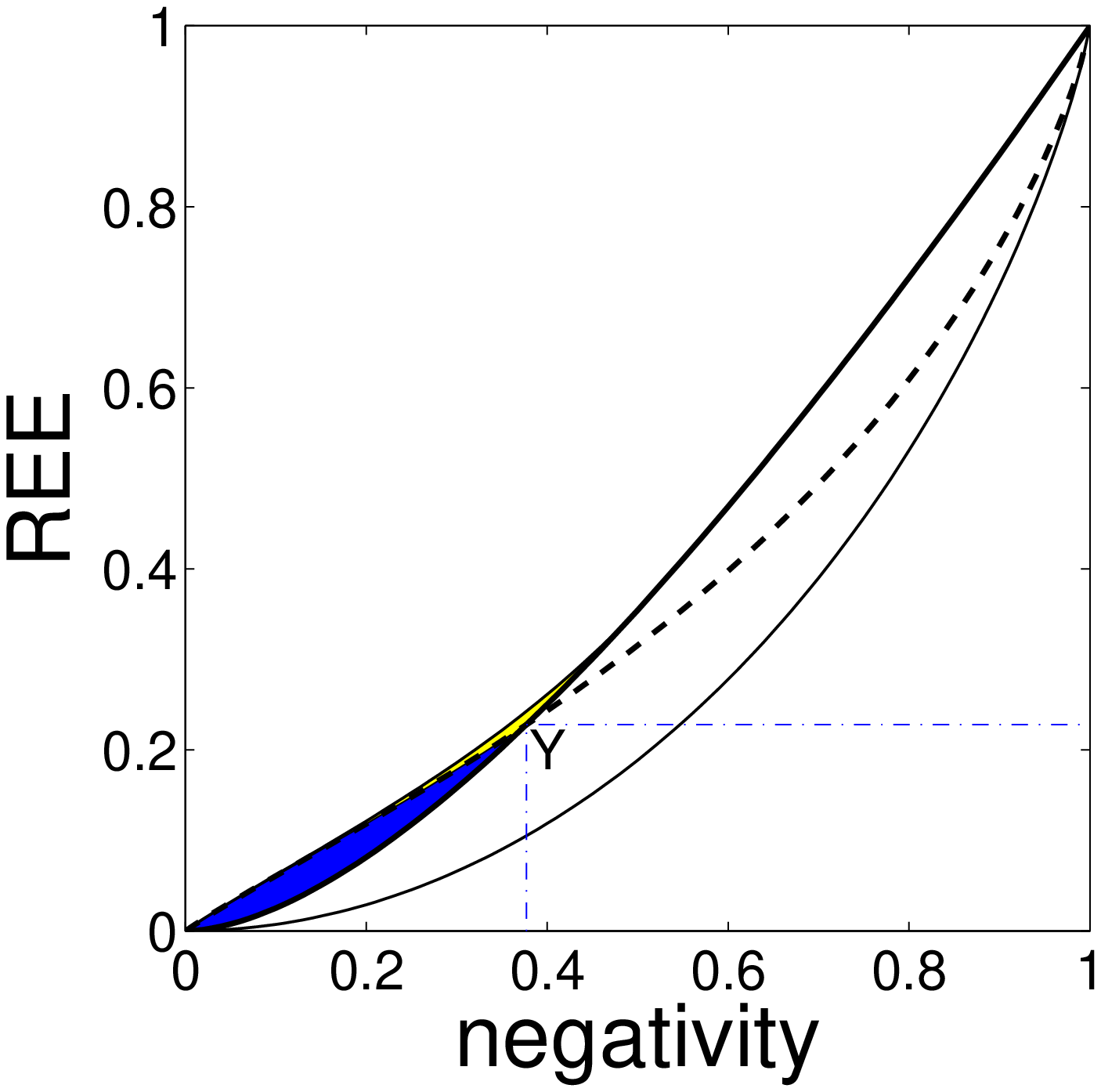}}
\vspace*{0cm} \caption{(Color online) The relative entropy of
entanglement (REE), $E_R(\hat \rho)$, as a function of the
negativity, $N(\hat \rho)$, for pure states (thick solid curve),
Horodecki states (dashed curve) and Bell diagonal states (think
solid curve). Blue and yellow regions correspond to mixed states
with the REE higher than that for pure states for a given value
the negativity. States in blue region are described in the text.}
\end{figure}

It can be shown analytically that pure states are the upper bound
for the negativity for a given value of the
concurrence~\cite{Verstraete01a}, as shown in Fig.~3(a), and the
upper bound for the REE as a function of the
concurrence~\cite{Vedral98}, as presented in Fig.~3(b). Similar
conclusions can be drawn for, e.g., the nonlocality for a given
value of the concurrence [see Fig.~3(c)], and the nonlocality as a
function of the negativity.

Thus, it is reasonable to conjecture that pure states are the
upper bound also for the REE, e.g., for a given value of the
negativity. But it was shown in Refs.~\cite{H7,H8} that this
conjecture is wrong [see Fig.~3(e)]. This property can be
demonstrated analytically on the example of, e.g., the Horodecki
state~\cite{Horodecki-review} defined as a mixture of the
maximally entangled state [e.g., the singlet state
$|\Psi_1\rangle$] and a separable state orthogonal to it (e.g.,
$|00\rangle$):
\begin{eqnarray}
\hat\rho ^{(H)}=p|\Psi_1\rangle \langle \Psi_1|+(1-p)|00\rangle
\langle 00|, \label{N42}
\end{eqnarray}
where $p\in\langle 0,1 \rangle$. The negativity and REE for the
Horodecki state are equal to
\begin{eqnarray}
N(\hat\rho ^{(H)})&=&\sqrt{ (1-p)^{2}+p^{2}}-(1-p), \label{N43}\\
E_R(\hat\rho ^{(H)}) &\equiv& E_R^{(H)}(N) = 2h(1-p/2)-h(p)-p,
\label{N44}
\end{eqnarray}
respectively, where $p=\sqrt{2N(1+N)}-N$ and $h(x)$ is binary
entropy. By comparing the REEs for Horodecki's state and for pure
states, it can be shown that~\cite{H7,H8}:
\begin{eqnarray}
 E_R^{(H)}(N) > E_R^{(P)}(N)& \quad &{\rm for}\; 0<N<N_Y,
 \label{N201}\\
 E_R^{(H)}(N) < E_R^{(P)}(N)& \quad &{\rm for}\; N_Y<N<1,
 \label{N202}
\end{eqnarray}
where $N_Y=0.3770\cdots$ and $E_R^{(H)}(N_Y) = E_R^{(P)}(N_Y)
=0.2279\cdots$, which corresponds to point $Y$ in Fig.~4. These
inequalities were shown analytically by expending $E_R^{(H)}(N)$
and $E_R^{(P)}(N)$ in power series of $N=\epsilon$
($N=1-\epsilon$) for values close to 0 (1). Moreover, mixed states
corresponding to blue region in Fig.~4, for which the inequality
in Eq.~(\ref{N201}) holds, can be obtained by mixing the Horodecki
state $\hat\rho_H$ with a separable state $\hat\rho^{(H)}_{\rm
css}$ closest to $\hat\rho_H$~\cite{H7}:
\begin{eqnarray}
\hat\rho^{(H')}(p,N)= (1-x)\hat\rho^{(H)}+x\hat\rho^{(H)}_{\rm
css}, \label{N101}
\end{eqnarray}
where $N\in\langle 0,1 \rangle$,
$p\in\langle\sqrt{2N(1+N)}-N,1\rangle$ and
$x=[(N+p)^2-2N(1+N)]/[p^2(1+N)]$. The closest separable state
$\hat\rho^{(H)}_{\rm css}$ is given by ($q=p/2$):
\begin{eqnarray}
\hat\rho^{(H)}_{\rm css}(p) &=&
{q}(1-{q})\sum_{j,k=0}^1(-1)^{j-k}|j,1-j\rangle \langle k,1-k|
+(1-{q})^{2}|00\rangle \langle 00|+{q}^{2}|11\rangle \langle 11|.
\label{N102}
\end{eqnarray}
By applying Vedral-Plenio's theorem~\cite{Vedral98}, the REE can
be found as follows~\cite{H7}:
\begin{eqnarray}
E_R(\hat\rho^{(H')}) \equiv E_R^{(H')}(p,N) &=& q^2 x\lg
x+2qy_1\lg \Big(\frac{y_1}{1-q}\Big) + y_2\lg
\Big(\frac{y_2}{(1-q)^2}\Big), \label{N103}
\end{eqnarray}
where $y_1=1-qx$ and $y_2=1-2q+q^2x$. With this choice of $x$,
parameter $N$ is just the negativity of $\hat\rho^{(H')}(p,N)$.
States corresponding to blue region in Fig.~4 can be obtained as
special cases of state $\hat\rho^{(H')}(p,N)$ for $N$ in the range
$0<N<N_Y$ and proper values of $p$. Thus, it is seen that there
are mixed states for which the REE is greater than that for pure
states at least in the range $N\in (0,N_Y)$. Later, in
Ref.~\cite{H8}, it was shown that the generalized Horodecki states
exhibit this property in slightly larger range as shown by yellow
region in Fig.~4. There is some evidence~\cite{H8} that the upper
bound of the REE as a function of the negativity is likely to be
given by these states.

Recently, we also analytically demonstrated~\cite{Horst} that the
entanglement REE for a given nonlocality for mixed states exceeds
that for pure states [see Fig.~3(f)]. Moreover, this effect occurs
in the larger range of abscissa values in comparison to the
dependence of the REE on the negativity, as seen by comparing
Figs.~3(e) and 3(f).

\section{Conclusion}

In this short review, we presented a few intriguing properties of
some standard entanglement measures for two qubits. Our examples
include a comparison of the negativity corresponding to the
Peres-Horodecki criterion~\cite{Peres,Horodecki96}, the Wootters
concurrence~\cite{Wootters}, and the relative entropy of
entanglement of Vedral \etal~\cite{Vedral97a}. Moreover, the
predictions of these measures were also compared with the
Horodecki measure~\cite{Horodecki95} of the violation of Bell's
inequality, referred here to as ``nonlocality''.

We discussed the following three counterintuitive properties of
entanglement measures: (i) entangled states {\em cannot} be
ordered uniquely with the entanglement measures, which also
implies that (ii) fragility or robustness of entanglement of
dissipative systems {\em cannot} be uniquely classified by
entanglement measures, and (iii) there are two-qubit mixed states,
which are more entangled (according to the REE) than pure states
for a given negativity or nonlocality.

It is well known that there is no unique entanglement measure for
mixed states. But the relativity of entanglement measures and its
implications are more counterintuitive. Our demonstration might
indicate that operational approaches to the quantum entanglement
problem are more meaningful rather than standard approaches based
formally-defined measures. We find the problem of defining
operational entanglement measures analogous to operational
approaches to the quantum phase problem~\footnote{Noh \etal~in
Ref.~\cite{Noh92} wrote: ``There has been a good deal of
discussion in the past of the most appropriate dynamical variable
to represent the phase of a quantum field, and many candidates
have been studied. Our analysis suggests that this question may
not have a general answer with respect to the measured phase
operators, because different measurement schemes lead to different
operators. As in many other quantum-mechanical problems, it seems
that questions about the value of a dynamical variable cannot be
divorced from the measurement process that generates the
ensemble.''} posed by Noh \etal~\cite{Noh91,Noh92}. The idea is to
define entanglement (or phase) measures in terms of what actually
is, or can be, measured.

We hope that the discussed problem of non-unique ordering of
states according to formally-defined entanglement measures can
stimulate investigations of operationally-defined measures
oriented for some specific experiments.

\noindent {\bf Acknowledgements} The work was supported by the
Polish Ministry of Science and Higher Education under Grant No.
N~N202 261938.

\appendix

\section{Notes on calculation of the REE}

The concurrence, negativity and nonlocality can be calculated
easily. By contrast, there has not yet been proposed an efficient
method to calculate the REE for arbitrary mixed states even in the
case of two qubits~\cite{Eisert}. Analytical formulas for the REE
are known only for some special sets of states with high symmetry
(see~\cite{Horodecki-review,H9} and references therein). Thus,
usually, numerical methods for calculating the REE have to be
applied~\cite{Vedral98,rehacek,doherty}.

It is a long-standing problem, posed by Eisert~\cite{Eisert}, of
obtaining an analytical compact formula for the REE for two
qubits. The problem is equivalent to finding the closest separable
state $\hat\rho_{\rm css}$ for a given entangled state $\hat\rho$.
In Ref.~\cite{H9}, a few arguments were given indicating that this
problem, probably, cannot be solved analytically for arbitrary
states. Nevertheless, there exists a solution to the inverse
problem of finding an analytical formula for $\hat \rho$ for a
given closest separable state $\hat\rho_{\rm css}$ as derived by
Ishizaka \etal~\cite{Ishizaka03,H9}.

The complexity of the problem can be explained (see,
e.g.,~\cite{Vedral98}) by virtue of Caratheodory's theorem, which
implies that any separable two-qubit state can be decomposed as
\begin{eqnarray}
\hat\rho_{\rm sep} = \sum_{j=1}^{16} p^2_j
|\psi_{j}^{(1)}\rangle\langle\psi_{j}^{(1)}| \otimes
|\psi_{j}^{(2)}\rangle\langle\psi_{j}^{(2)}|, \label{N21}
\end{eqnarray}
where the $k$th ($k=1,2$) qubit pure states can be parametrized,
e.g., as follows
\begin{equation}
|\psi_{j}^{(k)}\rangle=\cos\alpha_{j}^{(k)}|0\rangle+
\exp(i\eta_{j}^{(k)})\cos\alpha_{j}^{(k)}|1\rangle,
\end{equation}
and $p_j=\sin\phi_{j-1}\prod_{i=j}^{15}\cos\phi_i$ with
$\phi_0=\pi/2$. Thus, the minimalization of the quantum relative
entropy $S(\hat\rho ||\hat\rho_{\rm sep} )$, given by
Eq.~(\ref{N02}), with $\hat\rho_{\rm sep}$ described by
Eq.~(\ref{N21}), should be performed over $16\times 4 + 15 = 79$
real parameters. Usually (see, e.g.,
Refs.~\cite{Vedral98,rehacek}), gradient-type algorithms are
applied to perform the minimalization. \v{R}eh\'a\v{c}ek and
Hradil~\cite{rehacek} proposed a method resembling a state
reconstruction based on the maximum likelihood principle. Doherty
\etal~\cite{doherty} designed a hierarchy of more and more complex
operational separability criteria for which convex optimization
methods (known as semidefinite programs) can be applied
efficiently. One can also use an iterative method based on
Ishizaka formula~\cite{Ishizaka03,H9} for the closest entangled
state for a given separable state in order to find the closest
separable state for a given entangled state. Our algorithms for
calculating the REE are based either on the latter method or on a
simplex search method without using numerical or analytic
gradients.


\end{document}